\documentclass[aps,pra,preprint,groupedaddress,showpacs]{revtex4}
\usepackage[dvips]{epsfig}
\usepackage{subfigure}
\usepackage{color}
\usepackage{graphicx}

\begin{document}

\title{Power and momentum dependent soliton dynamics in lattices with longitudinal modulation}

\author{Panagiotis Papagiannis, Yannis Kominis and Kyriakos Hizanidis}
\affiliation{School of Electrical and Computer Engineering, National Technical University of Athens, 157 73 Athens, Greece}

\begin{abstract}
Soliton dynamics in a large variety of longitudinally modulated lattices are studied in terms of phase space analysis for an effective particle approach and direct numerical simulations. Complex soliton dynamics are shown to depend strongly on both their power/width and their initial momentum as well as on lattice parameters. A rish set of qualitatively distinct dynamical features of soliton propagation that have no counterpart in longitudinally uniform lattices is illustrated. This set includes cases of enhanced soliton mobility, dynamical switching, extended trapping in several transverse lattice periods, and quasiperiodic trapping, which are promising for soliton control applications. 
\end{abstract} 
\pacs{42.65.Sf, 42.65.Tg, 03.75.Lm, 05.45.Yv}
\maketitle

\section{Introduction}

The propagation of spatially localized waves in inhomogeneous nonlinear media is a subject of general  interest  that rises in many branches of physics such as light propagation in photonic lattices, matter-wave formation in Bose-Einstein condensates and solid state physics. Periodic modulation of the material properties in the form of a photonic lattice configuration has been shown to alter solitary wave formation and dynamics so that lattice solitons have features and properties with no counterpart in homogeneous media \cite{Chris}. The inhomogeneity of the medium can be either written in the form of waveguide arrays and optical networks or dynamically formed and leads to additional functionality with respect to wave control allowing for a variety of interesting applications. In such lattices stable solitons can be formed in specific positions determined by the lattice geometry. For the case of photonic structures transversely modulated by monochromatic (single wavenumber) variations of the linear or nonlinear refractive index stable solitons are always formed in the positions corresponding to the minima of the respective potential \cite{mono_static,OpEx_08}. In the case of polychromatic lattices the effective potential is different for solitons having different power and spatial width so that the number and the positions of stable solitons depend on their properties \cite{OpEx_08}. Therefore, the increased complexity of the medium modulation results in additional functionality of the photonic structure with respect to soliton discrimination.\

An additional degree of freedom in lattice modulation is related to the longitudinal periodic modulation of the properties of the medium. Such cases have been studied in the motion of charged-particle wave packets along periodic potentials in an a.c. electric field \cite{DuKe_86, Holthaus_92, DiSt_02} and in matter-wave dynamics in Bose-Einstein condensates under the influence of a time-periodic potential \cite{MaMa_06, PoAlOst_08, StLo_08}. In the context of light propagation in photonic lattices, previous studies include diffraction managed solitons \cite{dm} as well as Rabi-like oscillations and stimulated mode transitions with linear and nonlinear wave states in properly modulated waveguides and lattices \cite{rabi}. Longitudinally modulated lattices have been considered for soliton steering in dynamically induced lattices \cite{Kominis_steering, Tsopelas_steering, Other_dragging}. Dynamical localization and control has also been studied in longitudinally modulated waveguide arrays including periodically curved \cite{wgd_curved} waveguides and waveguides with periodically varying width \cite{wgd_varwidth} as well as lattices with longitudinally varying refractive index \cite{var_refractive}.\

In this work we investigate the effect of the a wide variety of different types of longitudinal lattice modulation on soliton dynamics and explore the additional functionality of the corresponding photonic structure in terms of soliton control based on both their power/width and momentum. Strong momentum dependence of soliton dynamics results from the fact that longitudinally modulated lattices actually carry momentum if seen as wave modulations of the medium \cite{Kominis_steering}. The two types of longitudinal modulations considered are amplitude and transverse wavenumber modulation along the propagation distance, being capable of describing several realistic configurations as well as providing fundamental understanding of the dynamical features related to even more general modulations. Soliton propagation in such inhomogeneous media is modeled by a NonLinear Schr\"{o}dinger (NLS) equation with transversely and longitudinally varying linear refractive index. Utilization of this continuous model allows for the study of soliton dynamical trapping and detrapping dynamics in contrast to discrete models where discrete solitons are locked at high powers to their input waveguides and are not allowed to travel sideways. \

In contrast to many previous studies where longitudinal and transverse length scales are well separated, our study focuses on cases where length scale interplay takes place. Therefore, the periods  of longitudinal lattice modulation are comparable to periods of soliton oscillation in a longitudinally homogeneous lattices. Resonances between these periodicities result in drastic modification of soliton dynamics and gives rise to a plethora of novel dynamical features depending on soliton power and momentum, including enhanced mobility, dynamical switching and trapping in extended areas of the lattice, periodic and quasiperiodic oscillations. An effective particle approach \cite{Kaup_Newell} is utilized in order to obtain a nonintegrable Hamiltonian system describing soliton "center of mass" motion. The complex dynamics of the system are analyzed in terms of Poincare surfaces of section providing a comprehensive illustration of soliton dynamical features. The remarkable agreement of direct numerical simulations with the phase space analysis of the effective particle dynamics suggest that the latter provides a useful tool for analyzing dynamics of different solitons propagating in lattices with different configurations as well as for designing lattices with desirable features.

\section{Model and effective particle approach}

The wave propagation in an inhomogeneous medium with Kerr-type nonlinearity is described by the perturbed NonLinear Schr\"{o}dinger (NLS) equation
\begin{equation}\label{eqn:LatticePDE}
i\frac{\partial u}{\partial z}+\frac{\partial^2 u}{\partial x^2} +2|u|^2u +\epsilon n(x,z)u=0
\end{equation}
where $z$ and $x$ are the normalized propagation distance and transverse coordinates respectively. We consider relatively weakly inhomogeneous media where the potential function $n(x,z)$ models longitudinally modulated lattices. $\epsilon$ is a small dimensionless parameter indicating the strength of the potential and $n(x,z)$ is periodic in $x$ and $z$. \

The lattice potential has a fundamental transverse ($x$) periodicity $(K_0^{-1})$ and we consider two types of longitudinal ($z$) modulations. The first one will be called AM modulation, as we modulate the amplitude of the potential function according to  
\begin{equation}\label{eqn:AMmod}  
n_{AM}(x,z)=[A_0+\alpha\sin(K_1 x+\phi)\sin(\Omega z)]\sin(K_0 x)
\end{equation}
In Figs. 1(a)-(c) we demonstrate three examples of such lattice patterns in the $(x,z)$ plane. For the second type of modulation, which we call WM, the transverse wavenumber of the  lattice varies periodically with $z$ as follows  
\begin{equation}\label{eqn:FMmod}  
n_{WM}(x,z)=A_0\sin[K_0 x+\alpha\sin(K_1 x+\phi)\sin(\Omega z)]
\end{equation}
and the corresponding patterns are shown in Figs. 1(d)-(f). 
$A_0$ is the strength of the unperturbed transverse lattice which is set to unity without loss of generality, and $\alpha$ can be seen as the relative strength of the AM or WM modulation. The wavenumber of the longitudinal modulation is $\Omega$ and the secondary transverse wavenumber $K_{1}$ corresponds to a nonuniform longitudinal modulation depending also on the transverse coordinate, thus allowing for the description of a very wide variety of lattice patterns: We are able to model modulations where the positions of the potential maxima and minima are either varying in-phase or out-of-phase as in Figs. 1(a)-(c) or are curved as in Figs. 1(d)-(f), with respect to $z$.

The unperturbed NLS ($\epsilon=0$) has a fundamental singe-soliton solution of the form 
\[
u(x,z)=\eta sech[\eta(x-x_0)]e^{i(\frac{v}{2}z+2\sigma)}
\]
where $\dot{x}_0=v$, $\dot{\sigma}=-v^2/8+\eta^2/2$. $\eta$ is the amplitude or the inverse width of the soliton solution, $x_0$ its center (called sometimes the center of mass due to the effective-particle analogy of the solitons), $v$  the velocity, and $\sigma$ the nonlinear phase shift.

The longitudinal evolution of the center $x_0$ under the lattice perturbation is obtained by applying the effective-particle method. According to this method we assume that the functional form and the properties (width, power) of the soliton are conserved in the case of the weakly perturbed NLS. This assumption has to be verified through numerical integration of Eq. (\ref{eqn:LatticePDE}) at least to a good approximation. The equation for $x_0$ is equivalent to an equation that describes the motion of a particle under the influence of an effective potential (periodic in our case) and is given by
\begin{equation}\label{eqn:EffPartODE}
m\ddot{x}_0=-\frac{\partial V_{eff}}{\partial x_0}
\end{equation}
where $m=\int{|u|^2dx}$ the integral of the non-dimensional soliton power, equivalent to the particle mass, and $V_{eff}(x_0)=2\int{n(x,z)|u(x;x_0)|^2dx}$ the aforementioned  effective potential determining soliton dynamics. By substituting the expressions of fundamental soliton and lattice potential Eq. (\ref{eqn:AMmod}) or Eq. (\ref{eqn:FMmod}), we obtain $m=2\eta$ and
\begin{eqnarray}\label{eqn:V_AM}
V_{eff}^{(AM)}&=&2A_0\pi\epsilon\frac{K_0\sin(K_0 x_0)}{\sinh(\pi K_0/2\eta)}+\pi\epsilon\alpha\sin(\Omega z)\nonumber\\
			                        &&\times\sum_{j=\pm}\frac{K_{j}\cos(K_{j} x_0+\phi)}{\sinh(\pi  K_{j}/2\eta)} 
\end{eqnarray}
with $K_{\pm}=K_1\pm K_0$, and
\begin{equation}\label{eqn:V_WM}
V_{eff}^{(WM)}=2A_0\pi\epsilon\sum_{m=0}^{\infty}J_m(\alpha\sin(\Omega z))\frac{K_m\cos(K_m x_0+\phi)}{\sinh(\pi K_m/2\eta)}
\end{equation}
with $K_{m}=mK_1+K_0$, for the two modulation types respectively.\

The soliton dynamics, as described by Eq. (\ref{eqn:EffPartODE}), with the above effective potentials are determined by a nonautonomous Hamiltonian system  
\begin{equation}
H=\frac{mv^2}{2}+V_{eff}(x,z)
\end{equation}
where $z$ is considered as "time" and $v=\dot{x_0}$ is the velocity of soliton center of mass.\

For the case of zero longitudinal modulation ($\alpha=0$) the system is integrable and soliton dynamics are completely determined by the form of the $z$-independent effective potential. In such case solitons are either trapped, oscillating between two maxima of the effective potential, or traveling along the transversely inhomogeneous medium. For trapped solitons the oscillation frequency varies from a maximum frequency 
\begin{equation}\label{eqn:FundamentalFreq}   
\omega_0=K_0\sqrt{2\epsilon A_0 \frac{\pi K_0/2\eta}{ \sinh(\pi K_0/2\eta)}}
\end{equation} 
corresponding to small harmonic oscillations around the minimum of the effective potential to a minimum zero frequency (infinite period) corresponding to an heteroclinic orbit connecting the unstable saddle points located at the maxima of the effective potential. The heteroclinic orbit is the separatrix between trapped and traveling solitons. Conditions for soliton trapping are determined by the initial soliton energy $H$, which depends on both soliton position and velocity, and occurs when $-\omega_0^2/K_0^2\leq 2\eta H<\omega_0^2/K_0^2$. Moreover, solitons located at the stable points (corresponding to the minima of the effective potential) can be detrapped if their initial velocities exceed a critical value 
\begin{equation}
v_{cr}=\pm2\sqrt{max(V_{eff})/m}=\pm 2\omega_0/K_0
\end{equation}
and travel at a direction determined by the sign of their initial velocity.\

The presence of an explicit $z$ dependence in the effective potential ($\alpha\neq0$) results in the nonintegrability of the Hamiltonian system which describes the soliton motion and allows for a plethora of qualitatively different soliton evolution scenarios. The corresponding richness and complexity of soliton dynamics opens a large range of possibilities for interesting applications where the underlying inhomogeneity results in advanced functionality of the medium. The nonintegrability results in the destruction of the heteroclinic orbit (separatrix), allowing for dynamical trapping and detrapping of solitons. Therefore, we can have conditions for enhanced soliton mobility: Solitons with small initial energy can travel through the lattice as well as hop between adjacent potential wells and dynamically be trapped in a much wider area, including several potential minima. Moreover, resonances between the frequencies of the internal unperturbed soliton motion and the $z$-modulation frequencies result in quasiperiodic trapping and symmetry breaking with respect to the velocity sign. It is worth mentioning that all these properties depend strongly on soliton characteristics, namely $\eta$, so that different solitons undergo qualitatively and quantitatively distinct dynamical evolution in the same inhomogeneous medium.

\section{Results and Discussion}

In the following, we consider relatively weakly modulated lattices being of interest in most applications and set $\epsilon=10^{-2}$ . For example, in typical optical lattices consisting of nonlinear material of AlGaAs type (refractive index $n_0 \simeq 3.32$ at $\lambda_0 \simeq 1.53 \mu m$), when the trasnverse dimension $x$ is normalized to $1 \div 3\mu m$, $\epsilon=10^{-2}$ corresponds to an actual refractive index contrast $\Delta n_0=10^{-4} \div 10^{-5}$ which is relevant for experimental configurations. For such cases a normalized propagation distance $z=500$, which is large enough in order to observe the dynamical features presented in the following, corresponds to an actual length of $13 \div 123 mm$. Moreover, as we show in following, weak lattices are characterized by complex but not completely chaotic soliton dynamics, allowing for controllable evolution features.   
It is expected that interesting soliton dynamics occur when the spatial scales of the system, namely transverse and longitudinal modulation periods and soliton width, become comparable, so that length scale competition takes place. As shown from Eq. (\ref{eqn:FundamentalFreq}), the maximum oscillation frequency of a trapped soliton in an unmodulated lattice depends strongly on the ratio $K_0/2\eta$, so that solitons with different $\eta$ can have different $\omega_0$ as depicted in Fig. 2, where it is shown that the value of $\omega_0$ saturates quickly to an upper bound $\omega_0 (\eta\rightarrow\infty)=\sqrt{2\epsilon A_0}K_0$ as the width $(\eta^{-1})$ of the soliton becomes comparable or smaller than the fundamental transverse period ($2\pi/K_0$) of the unmodulated lattice. The value of $\omega_0$ for each soliton and $K_0$ is crucial since by introducing the longitudinal modulation, we expect the frequencies $\Omega$ that have significant effect in the soliton behavior are those that fulfill a resonance condition with the unperturbed frequencies ($\alpha=0)$ ranging from 0 to $\omega_0$. These resonant interactions give rise to qualitative and quantitative different evolution of $(x_0,v)$. In order to investigate the new features the longitudinal modulation induces, we use the phase space representation of (\ref{eqn:EffPartODE}) on a Poincare surface of section stroboscopically produced with period $\Omega$. The phase space topology visualizes all the qualitative features of soliton dynamics and provides a significant amount of comprehensive information which is useful for categorizing and analyzing characteristic cases and conceptual design of potential applications.\

As shown in Figs.3 and 4, in both AM and WM lattices the phase space topology changes significantly from the corresponding case with no longitudinal modulation (depicted with solid curves). The topology depends strongly on both the characteristics of the lattice and the characteristics of the soliton ($\eta$), so that not only different lattices but also different solitons on the same lattice can have drastically different dynamical features. This is shown in different columns and rows of Figs. 3 and 4, for transverse lattice modulations of period $\Lambda_0=2\pi (K_0=1)$ and soliton parameter $\eta=0.5 \div 2.5$ corresponding to a FWHM value range $5.3 \div 1$ in normalized transverse dimensions. 
Before proceeding to the investigation of specific characteristic cases of soliton evolution we discuss the qualitative topological characteristics of phase spaces corresponding to different lattices and soliton power ($\eta$). As expected for a nonintegrable system, the typical Poincare surface of section consists of regular curves corresponding to perturbed tori of nonresonant quasiperiodic orbits, resonant islands around periodic closed orbits and densely filled chaotic regions related to complex motion. In all cases, the separatrix between bounded and unbounded motion has been destroyed and replaced by a chaotic region while the regions corresponding to bounded motion in longitudinally modulated lattices are downsized. The extent of these areas depends strongly on the amplitude of the effective potential which varies exponentially on the ratio $K_m/\eta$ as shown in Eqs. (\ref{eqn:V_AM}) and (\ref{eqn:V_WM}). Therefore, solitons having different power ($\eta$) have completely different trapping conditions in the same lattice, while soliton mobility can be drastically enhanced. In addition, changing the soliton power for a specific lattice configuration results in a drastic change on the frequency spectrum of the unperturbed soliton oscillations through $\omega_0(\eta)$. This leads to the possibility of fulfilling resonance conditions with the frequency of the longitudinal lattice modulation $\Omega$ and to the appearance of periodic orbits and surrounding resonant islands. Another qualitative differentiation between the phase spaces is the bifurcation of the stable center corresponding to the minimum of the effective potential in the unmodulated lattice to a saddle point seen, for example, in Fig. 3(c),(d) for $\eta=1.75, 2.5$ ($\Omega/\omega_0=1, 0.97$ correspondingly). In addition to the power dependence of the phase space structure, a common characteristic of all phase spaces for longitudinally modulated lattices is the symmetry breaking of the Poincare surface of section with respect to $v=0$ and $K_0x_0=-\pi/2$. 
This feature reveals a selectivity of the lattice on initial velocity (momentum) and displacement: Solitons with opposite velocities or symmetrically placed from the former minimum of the effective potential (which in the unmodulated case would remain in the same orbit) they undergo now qualitatively distinct evolution. In such cases one of the solitons can be trapped while the other is detrapped. The velocity selectivity is a direct consequence of the momentum that is incorporated in the lattice pattern due to the biperiodic lattice potential and even though is met in both AM and WM lattices it seems to be more prominent in the later as seen from the Poincare surfaces of section. The dependence of motion on its initial displacement is related to the fact that the local minima of the transverse lattice profile changes periodically with $z$. \

Having discussed the topological features of phase spaces corresponding to different combinations of lattice configurations and soliton powers ($\eta$), we now show specific characteristic cases of soliton motion having qualitatively distinct properties and being promising for potential applications. Although similar cases can be met in many different cases, we focus our analysis on the cases depicted in Figs. 3(c) and 4(h) with parts of them illustrated in more detail in Figs. 5(a) and (b). In the following figures white thick lines and black dashed lines depict soliton center motion as obtained from numerical simulation of the original perturbed NLS Eq. (\ref{eqn:LatticePDE}) and the effective particle approach respectively, showing a remarkable agreement. The thick solid lines depict numerical simulations for the case of an unmodulated lattice for comparison.\

In Fig. 6(a) we illustrate soliton propagation for an initial center position and velocity corresponding to point (i) of Fig. 5(a). It is shown that soliton undergoes a complex evolution being dynamically trapped between several transverse lattice periods, in contrast to the case of same initial conditions in an unmodulated lattice. Soliton mobility is even more pronounced in the case shown in Fig. 6(b) where complete dynamical detrapping takes place allowing for a soliton to travel across the lattice. The latter is the typical case of soliton evolution for initial conditions located outside the region occupied by regular orbits in the corresponding phase spaces. It is worth mentioning that the degree of mobility enhancement is not the same for different solitons in the same lattice as it can be seen by the comparison of the size of the regular area in Figs. 3(a)-(d). Initial conditions located on the regular or island curves of the phase space, as point (iii) in Fig. 5(a), correspond to quasiperiodic soliton oscillations as shown in Fig. 6(c). In Fig. 6(d) the soliton evolution for an initial condition close to the saddle point (iv) of Fig. 5(a) is depicted exhibiting an unstable (hyperbolic) type of periodic orbit.\

A very interesting evolution scenario is illustrated in Fig. 7(a) for initial conditions depicted by points (i) and (ii) of Fig. 5(b), corresponding to solitons having the same initial positions but opposite initial velocities: The soliton with the positive velocity remains trapped and periodically oscillating (since it corresponds to a center of a resonant island - exact resonance) while the soliton with the negative velocity undergoes a dynamical switching to a neighbor lattice position where undergoes persistent trapping. This type of feature can be considered for promising power and velocity dependent soliton switching applications. In the same fashion, the effect of symmetry breaking with respect to $v=0$ can lead to velocity sign dependent soliton trapping or traveling across the lattice, as shown in Fig. 7(b). A trapped soliton propagation having the form of a beat is depicted in Fig. 7(c) for an initial condition corresponding to a resonant island.\          

As a final case we consider soliton dynamical trapping within an extended area including many transverse periods of the lattice in a persistent periodic fashion, as shown in Figs. 8(b) and (c). The initial conditions leading to such evolution correspond to two families of interconnected resonant islands (1) and (2) shown in Fig. 8(a). These islands are located outside the separatrix of the respective unmodulated lattice showing that outside but close to the separatrix the longitudinal modulation can induce interesting persistent dynamical trapping for initial conditions for which traveling solitons are expected in the unmodulated lattice. In accordance with previous cases, the lattice possesses a selectivity property with respect to initial soliton velocity direction as we have not a symmetry with respect to the initial velocity axis. The type of evolution depicted in Figs. 8(b) and (c) is interesting for power and momentum dependent multi-port soliton switching applications, where different input/output ports correspond to different potential minima.

\section{Conclusions}
We have studied soliton dynamics in a large variety of longitudinally modulated lattices in terms of direct numerical simulations as well as phase space analysis for an effective particle approach. The remarkable agreement of the results suggest that the effective particle approach and the phase space analysis with the utilization of Poincare surfaces of section provides a useful tool for studying complex soliton dynamics in such lattices as well as analyzing and/or designing lattices having desirable properties. It is shown that soliton dynamics depend strongly on their power and the corresponding spatial width through its relation with the transverse lattice period as well as on both magnitude and direction of their initial velocity. A large variety of qualitatively distinct dynamical features of soliton propagation have been shown that have no counterpart in longitudinally uniform lattices. Therefore, cases of enhanced soliton mobility, dynamical switching and trapping in several transverse lattice periods as well as quasiperiodic and periodic trapping have been shown, suggesting that the corresponding complexity of the effective particle phase space gives rise to a plethora of dynamical features which are promising for applications.

\begin{figure}[p]
 \begin{center}
    \scalebox{0.4}{\includegraphics[]{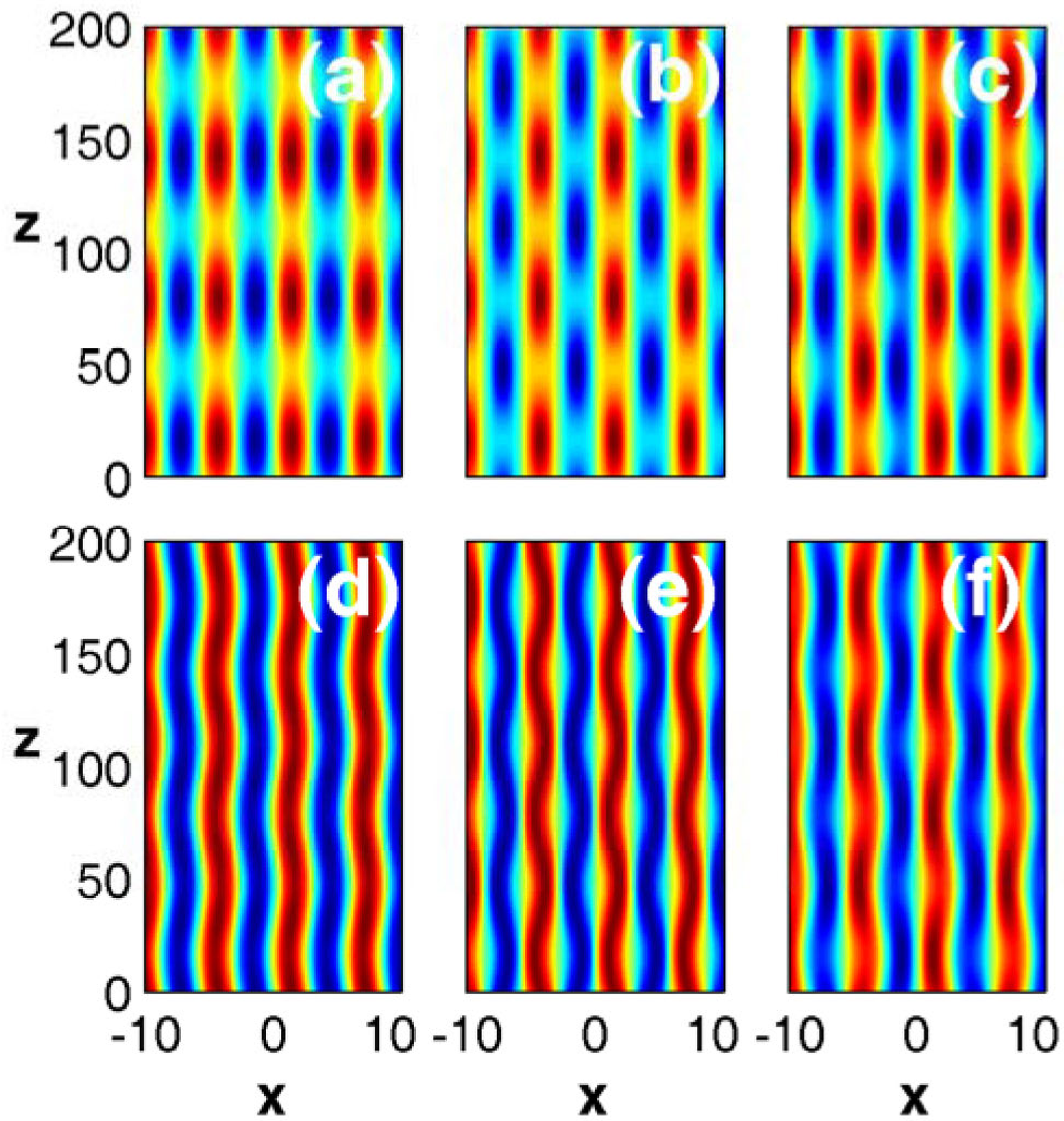}}
     \end{center}
     \caption{(Color online) Lattice patterns in ($x,z$) plane for AM (a-c)  and  WM (d-f) modulation. $K_0=1$ and $\Omega=0.1$ in all cases and  $K_1=0,\phi=\pi/2$ (a,d), $K_1=1,\phi=0$ (b,e), $K_1=1/2,\phi=0$ (c,f). Blue (light gray) colored areas indicate negative values and contain the minima of the lattice potential while red (colored areas indicate positive values where the corresponding maxima are located.}
\end{figure}
 
\begin{figure}[p]
   \begin{center}
       \scalebox{0.5}{\includegraphics{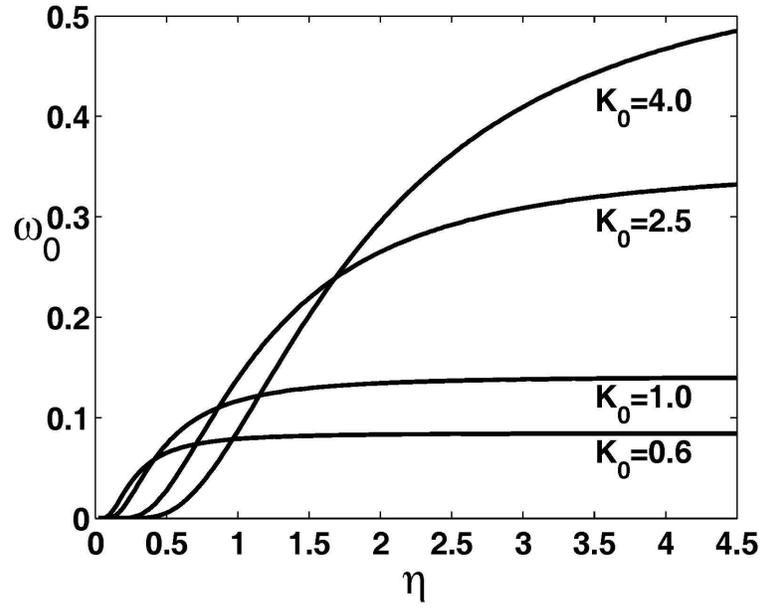}}
   \end{center}
   \caption{The spatial frequency of trapped soliton small oscillations $\omega_0$ as a function of inverse width $\eta$ for different values of the transverse wavenumber $K_0$ in an unmodulated lattice.}
\end{figure}

\begin{figure}[p]
  \begin{center}
     \scalebox{0.5}{\includegraphics[]{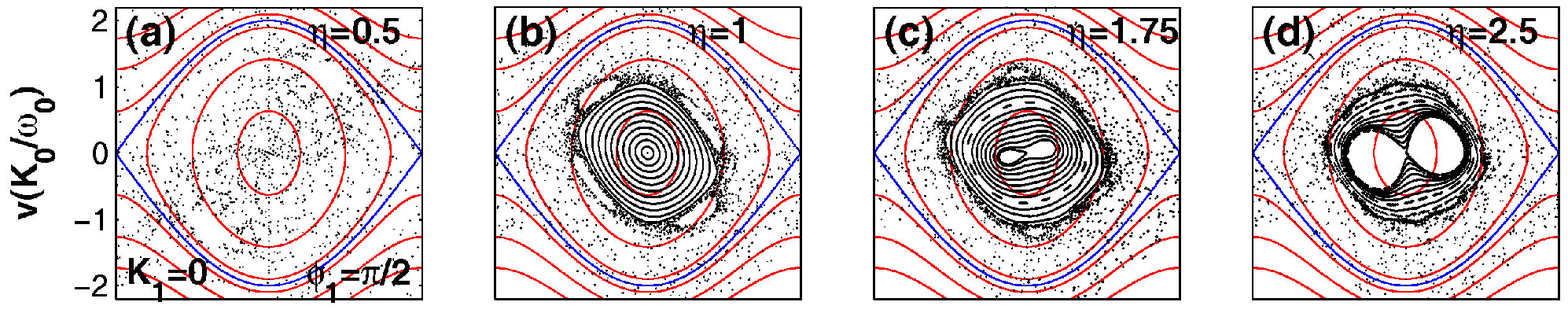}}
      \scalebox{0.5}{\includegraphics[]{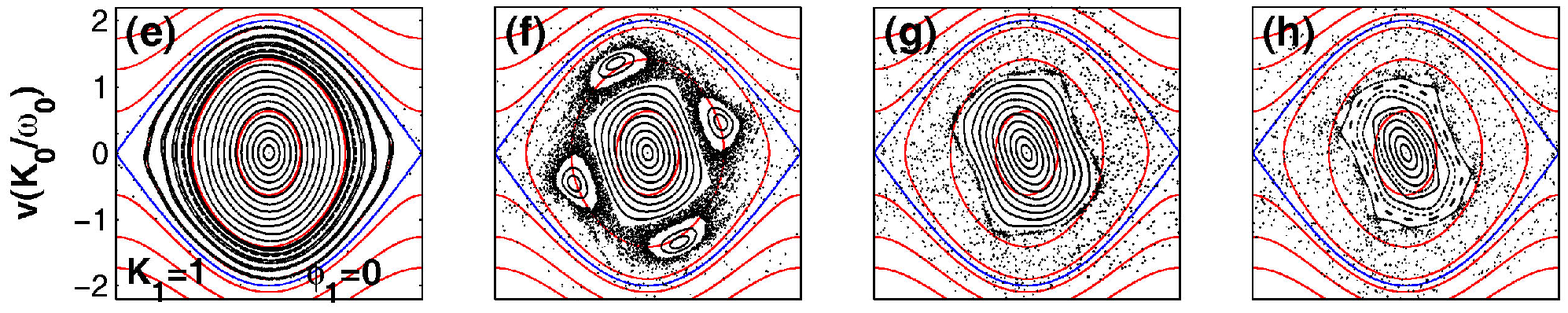}}
       \scalebox{0.5}{\includegraphics[]{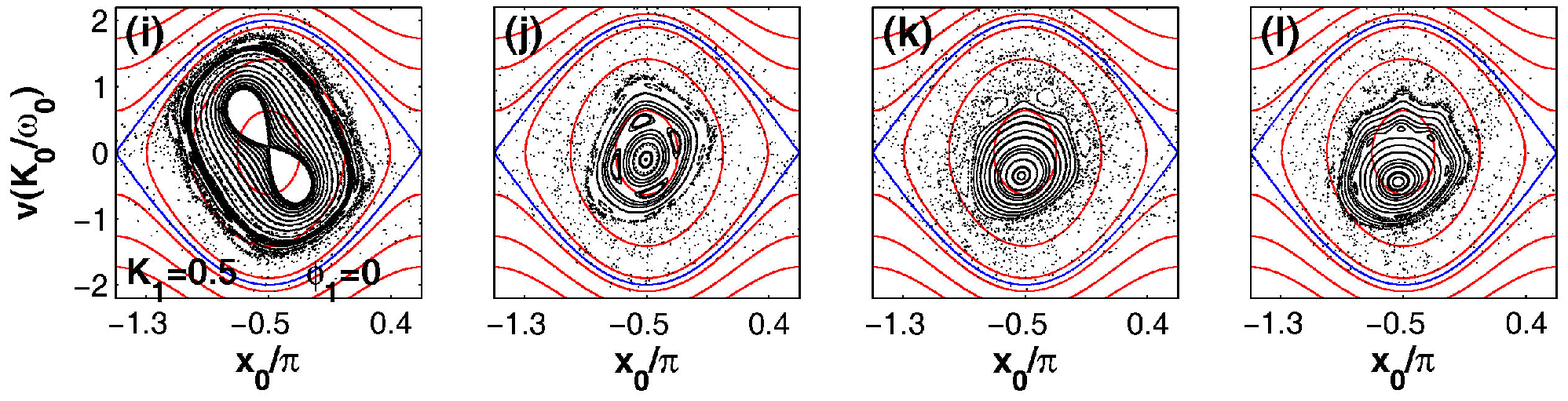}}
   \end{center}
      \caption{(Color online) Poincare surfaces of section of the effective-particle system for AM lattice modulations. Each row corresponds to the same lattice pattern (Figs.1(a)-(c)) and each column to the same soliton width with values $\eta=0.5, 1, 1.75,2$ from left to right . In all cases $K_0=1$. From top row to bottom, we have $\Omega=0.1325$, $\Omega=0.4$, $\Omega=0.15$ correspondingly. All surfaces of section are superimposed on the corresponding phase space without longitudinal modulation (continuous, red curves). The separatrix between the trapped and untrapped soliton motion in the unmodulated lattice (blue curve) is also shown.}
\end{figure}

\begin{figure}[p]
  \begin{center}
     \scalebox{0.5}{\includegraphics[]{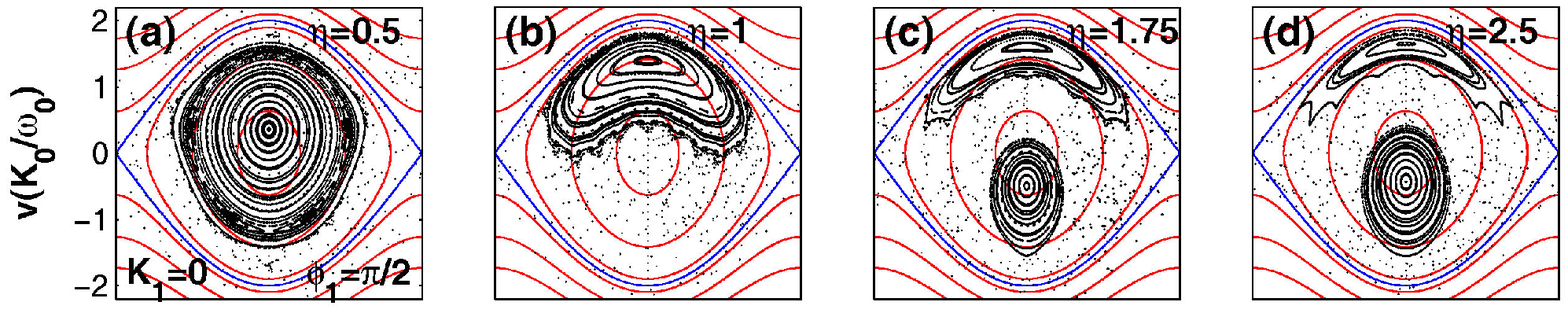}}
     \scalebox{0.5}{\includegraphics[]{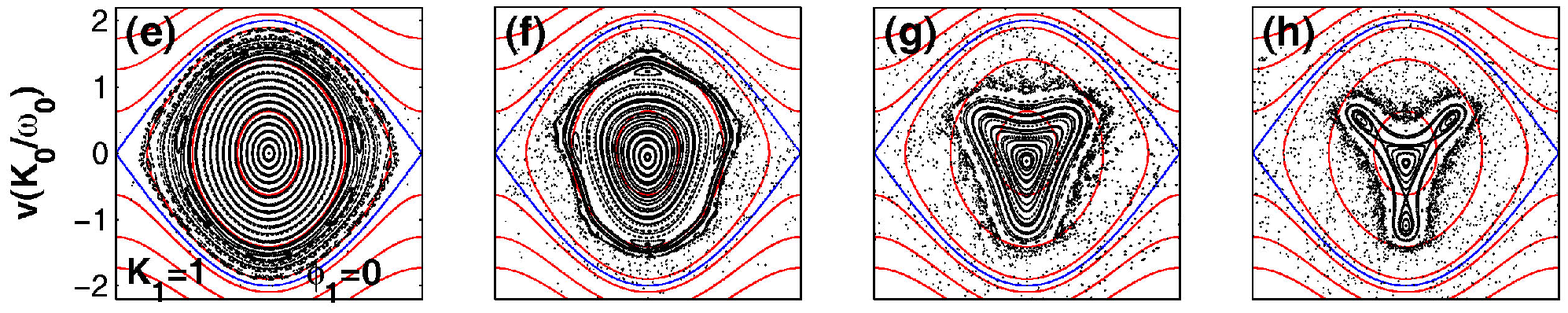}}
     \scalebox{0.5}{\includegraphics[]{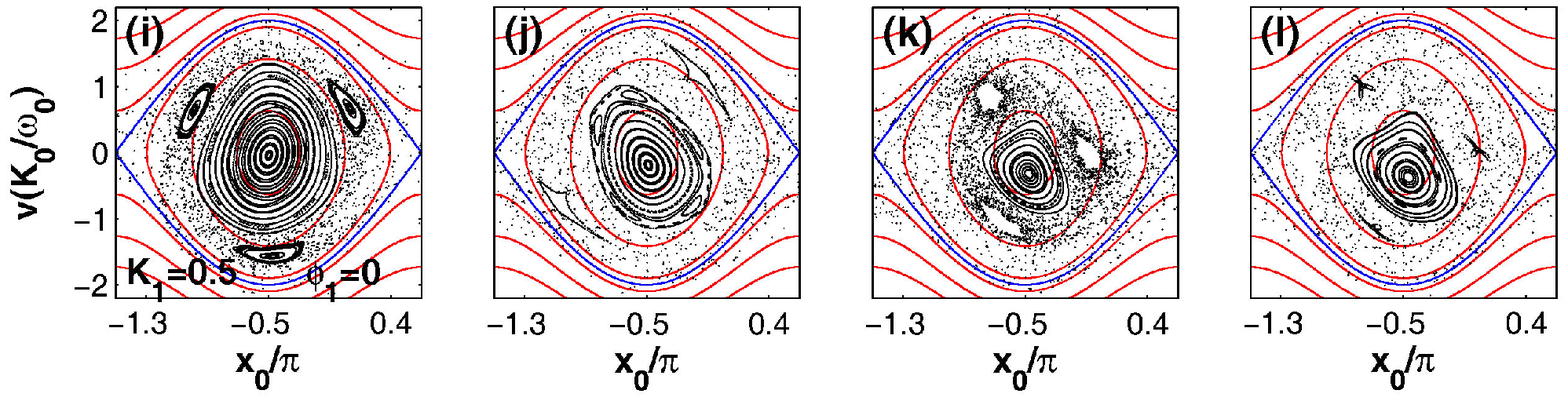}}
   \end{center}
      \caption{(Color online) Poincare surfaces of sections of the effective-particle system for WM lattice modulations. Each row corresponds to the same lattice pattern (Figs.1(d)-(f)) and each column to the same soliton width with values $\eta=0.5, 1, 1.75,2$ from left to right . From top row to bottom, we have $\Omega=0.1$, $\Omega=0.4$, $\Omega=0.175$ correspondingly. All surfaces of section are superimposed on the corresponding phase space without longitudinal modulation (continuous, red curves). The separatrix between the trapped and untrapped soliton motion in the unmodulated lattice (blue curve) is also shown.} 
\end{figure}

\begin{figure}[p]
    \begin{center}
     \subfigure[]{\scalebox{0.4}{\includegraphics{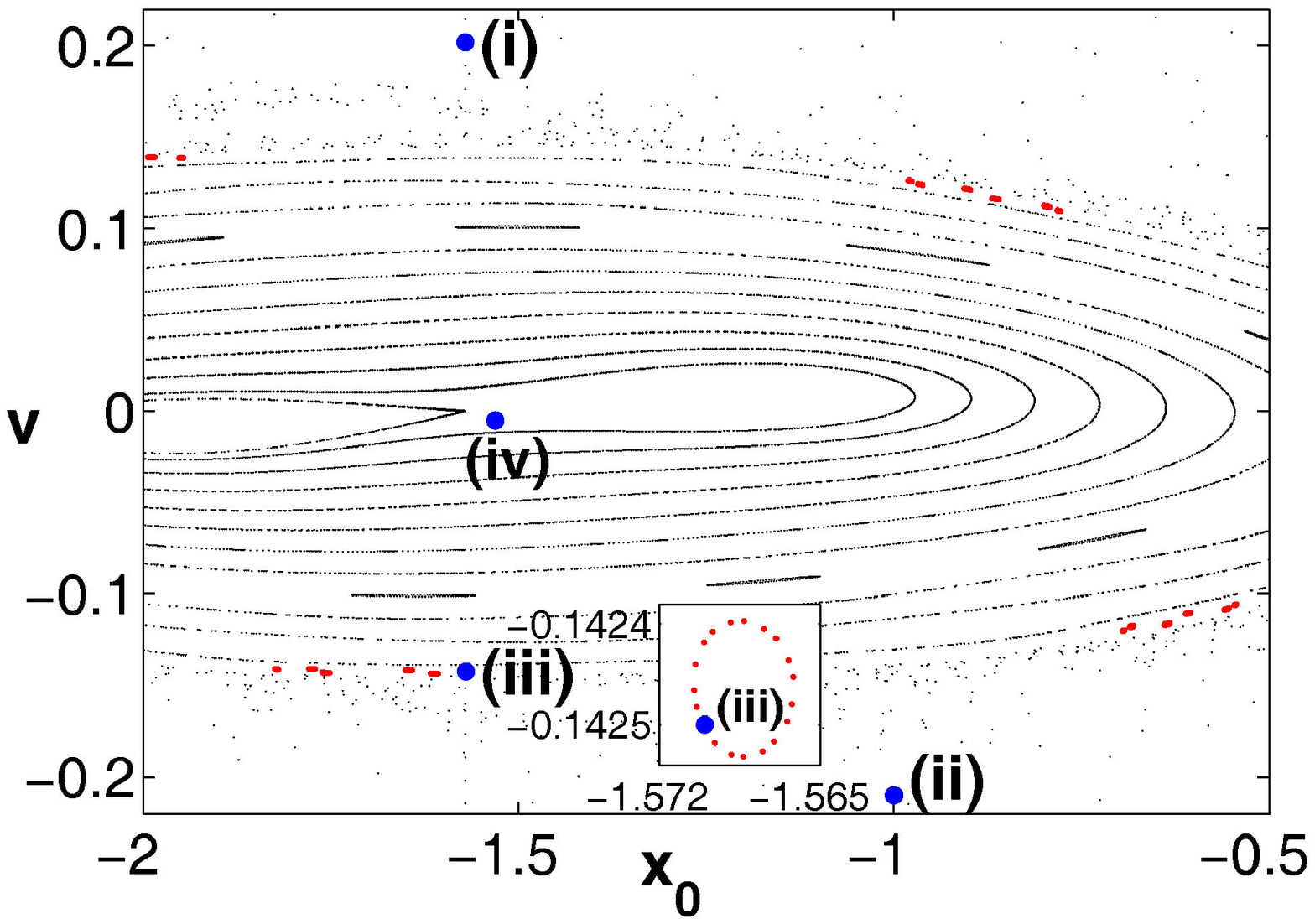}}}
     \subfigure[]{\scalebox{0.4}{\includegraphics{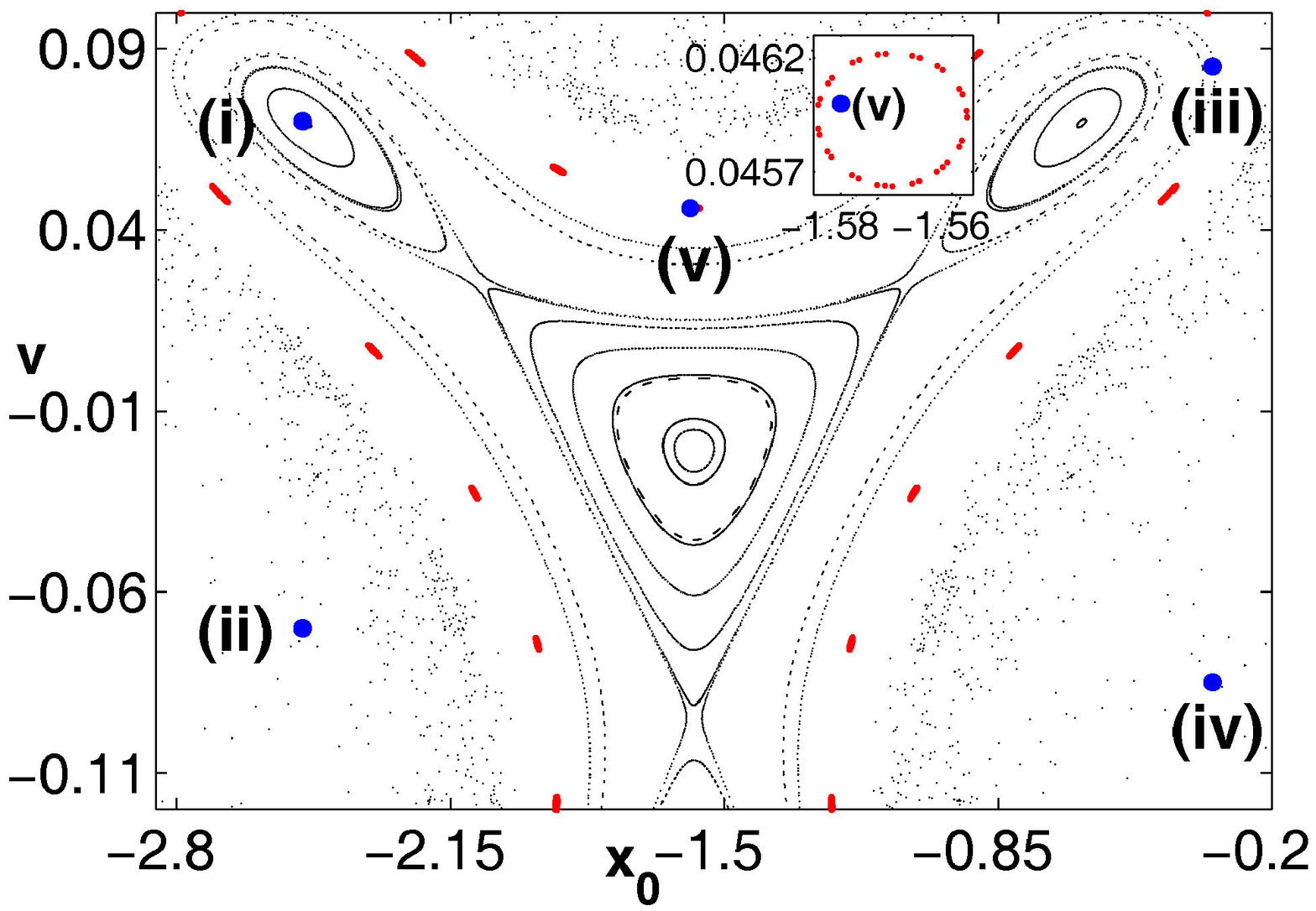}}}
       \caption{(Color online) (a) Detail of the Poincare surface of section from Fig.3(c) (AM modulation). The values of the initial parameters $(x_0,u)$ at the points depicted in figure are $(i)(-\pi/2,0.202)$, $(ii)(-1.57,-0.1425)$, $(iii)(-1,-0.21)$, $(iv)(-1.5308,-0.005)$.(b) Detail of the Poincare surface of section from Fig.4(h), (WM modulation). The values of the initial parameters $(x_0,u)$ at the points depicted in figure are $(i),(ii)(-2.5,\pm 0.07)$, $(iii),(iv)(-0.34,\pm 0.085)$, $(v)(-1.58,0.046)$.}
   \end{center}
\end{figure}

\begin{figure}[p]
    \begin{center}
     \subfigure[]{\scalebox{0.35}{\includegraphics{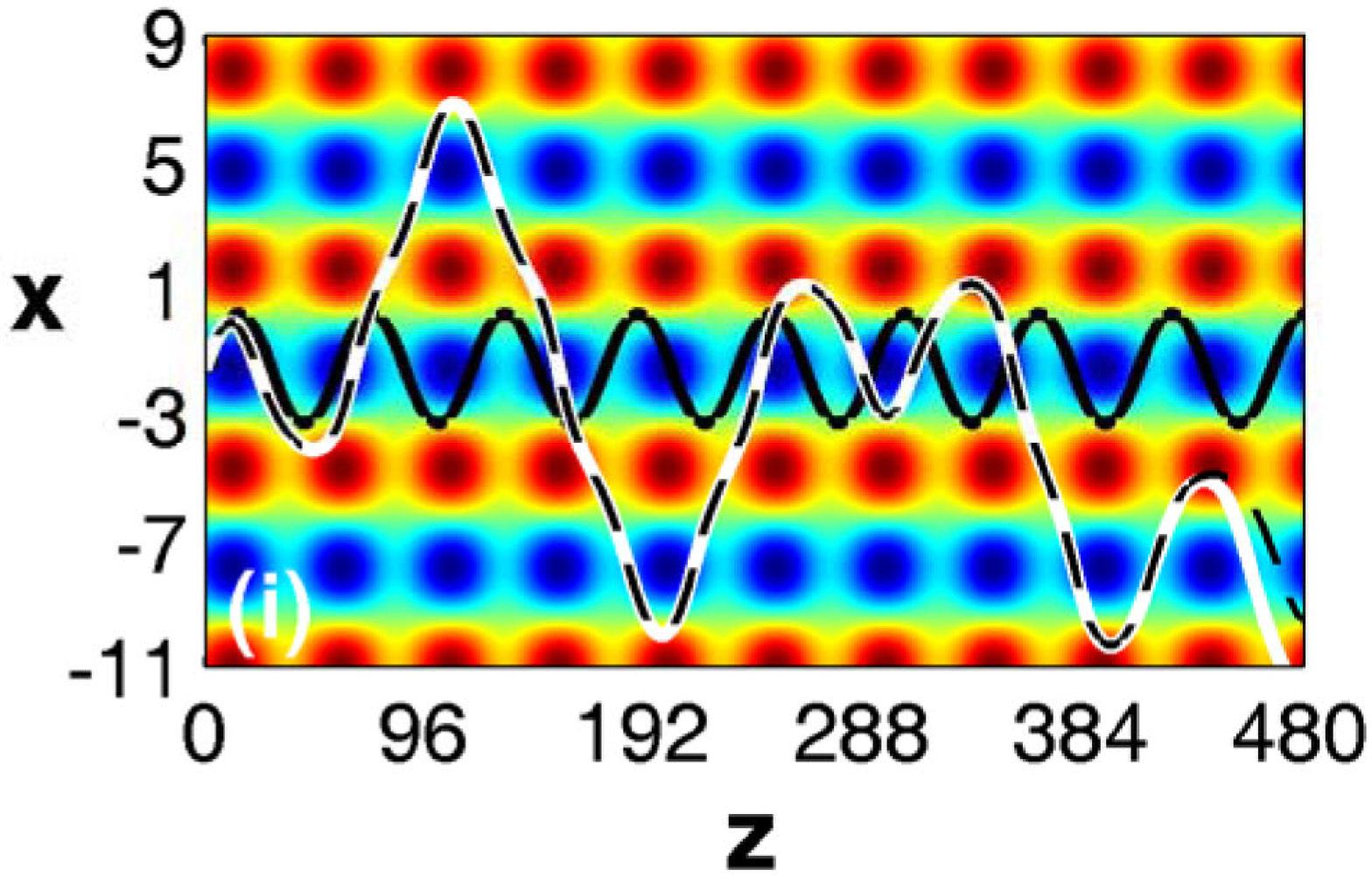}}}
     \subfigure[]{\scalebox{0.35}{\includegraphics{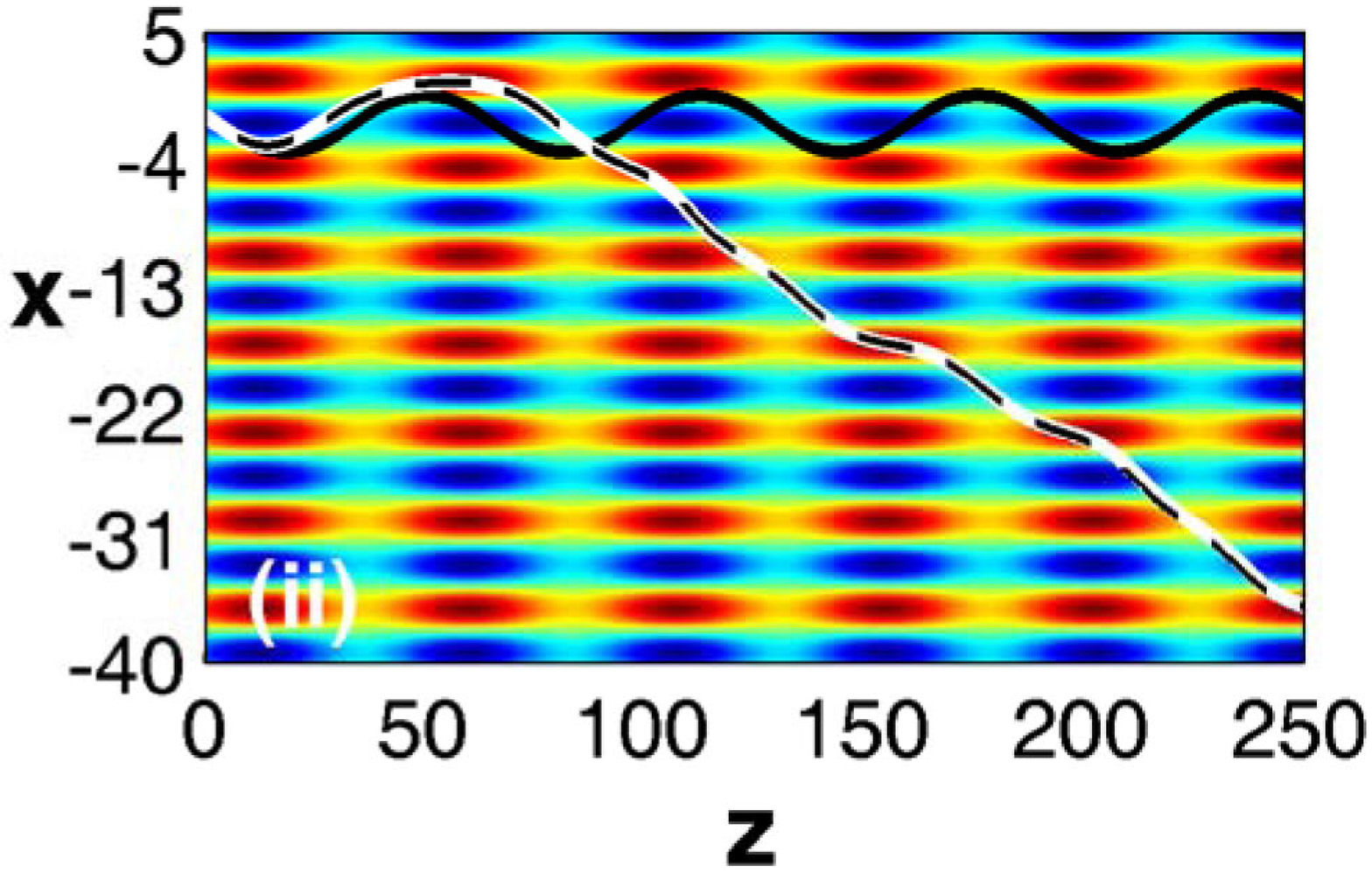}}}
     \subfigure[]{\scalebox{0.35}{\includegraphics{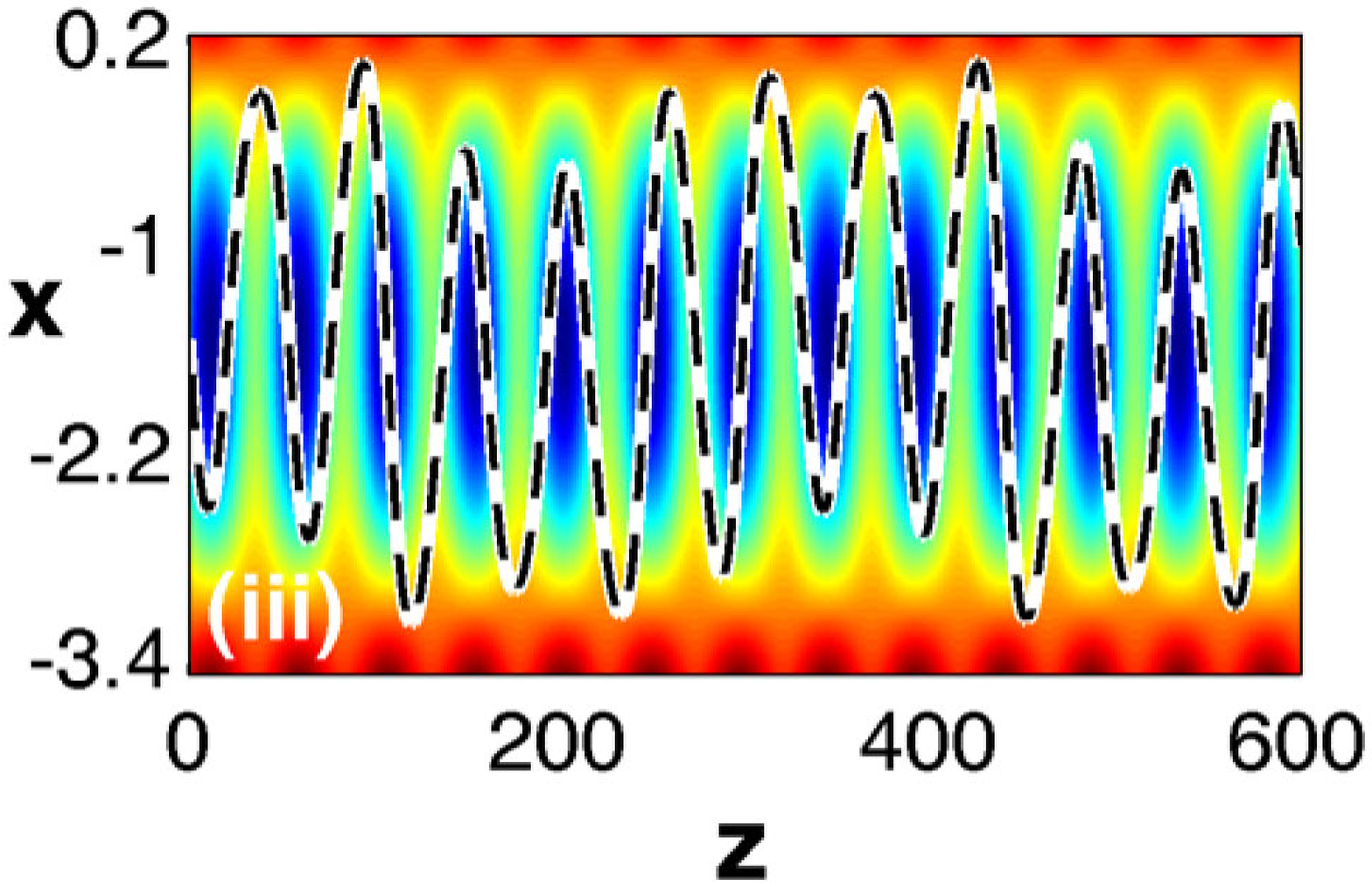}}}
     \subfigure[]{\scalebox{0.35}{\includegraphics{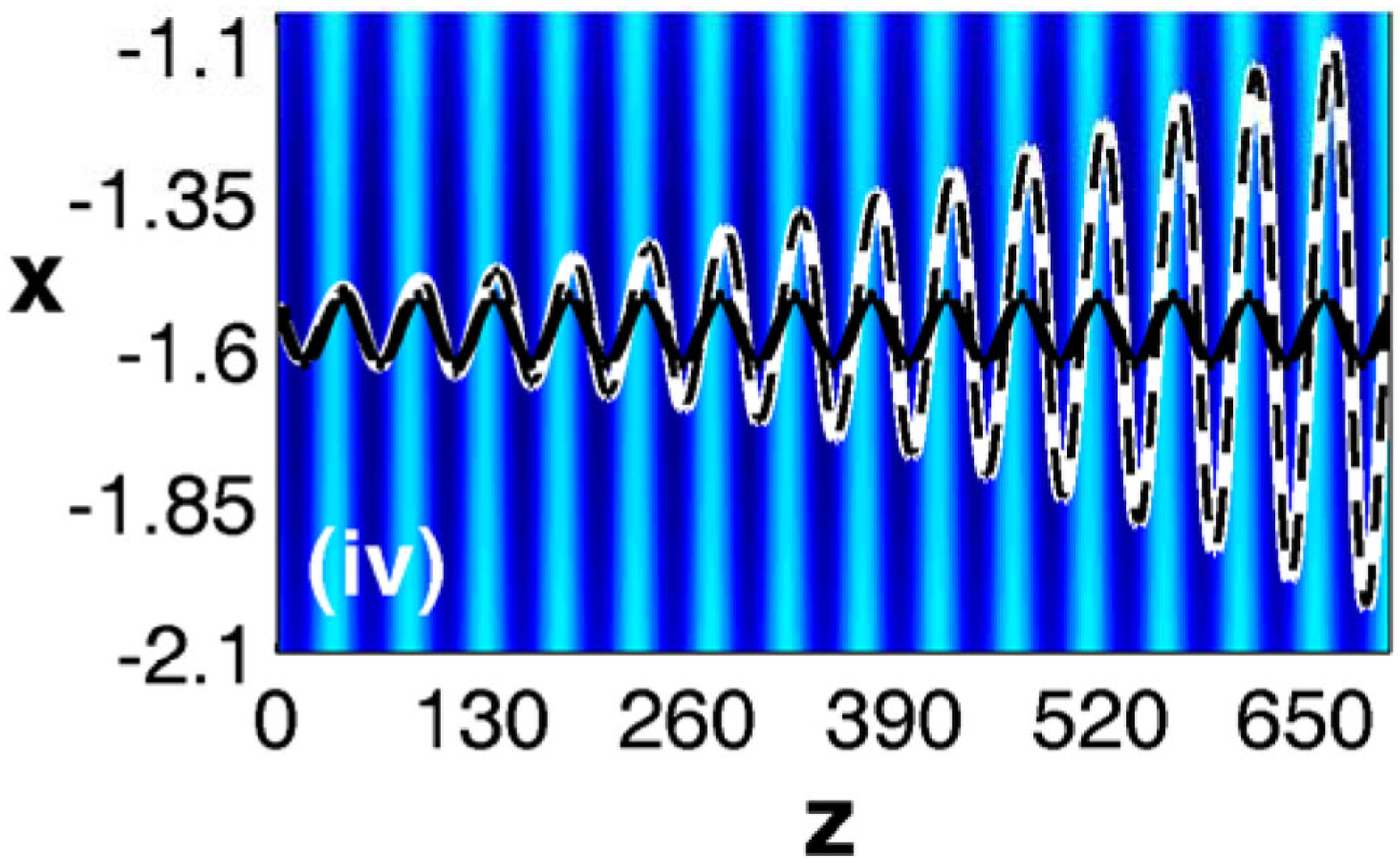}}}
       \caption{(Color online) Evolution of the soliton center in an AM lattice pattern with initial conditions taken from the corresponding points $(i)-(iv)$ of Fig. 5(a). (a) point $(i)$: enhanced mobility with dynamic trapping, (b) point $(ii)$: enhanced mobility with complete detrapping (c) point $(iii)$: quasiperiodic oscillations and (d) point $(iv)$: hyperbolic periodic oscillations. Thick white curves correspond to results from direct numerical simulations, dashed black curves from the effective particle model and solid black ones from direct simulations for the unmodulated lattice. The corresponding lattice pattern is shown in the background.}
   \end{center}
\end{figure}

\begin{figure}[p]
    \begin{center}
     \subfigure[]{\scalebox{0.35}{\includegraphics{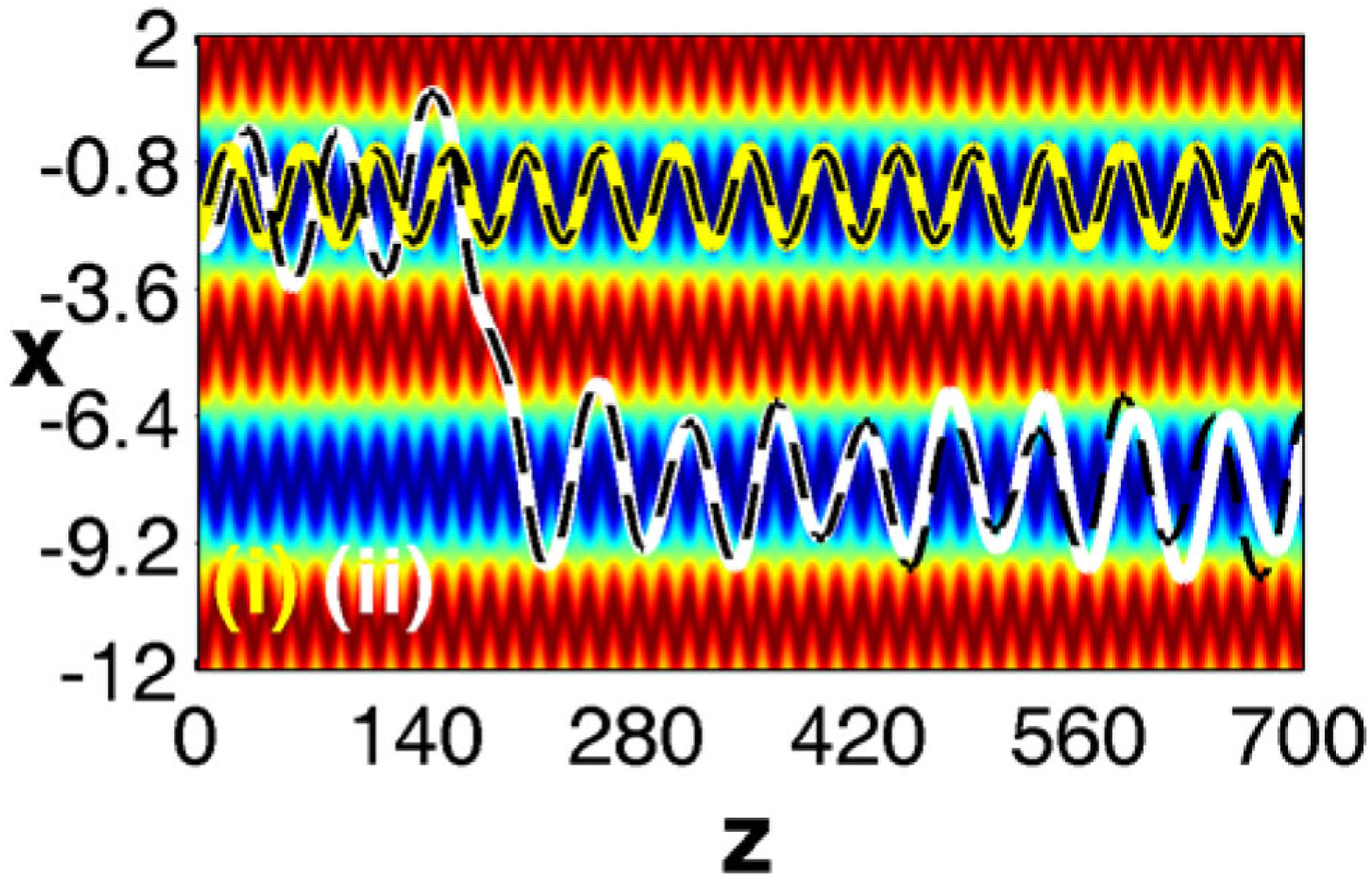}}}
     \subfigure[]{\scalebox{0.35}{\includegraphics{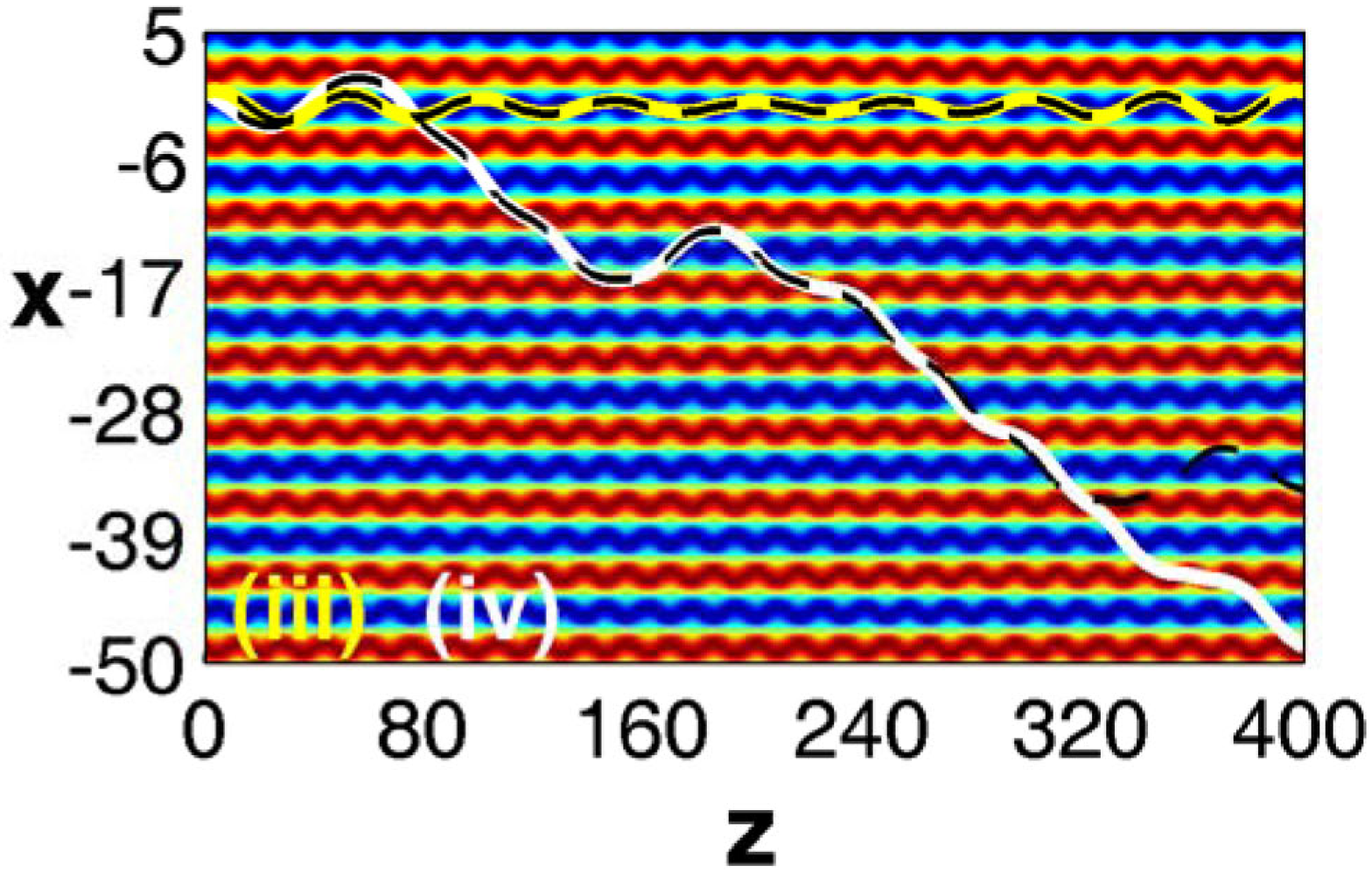}}}
     \subfigure[]{\scalebox{0.35}{\includegraphics{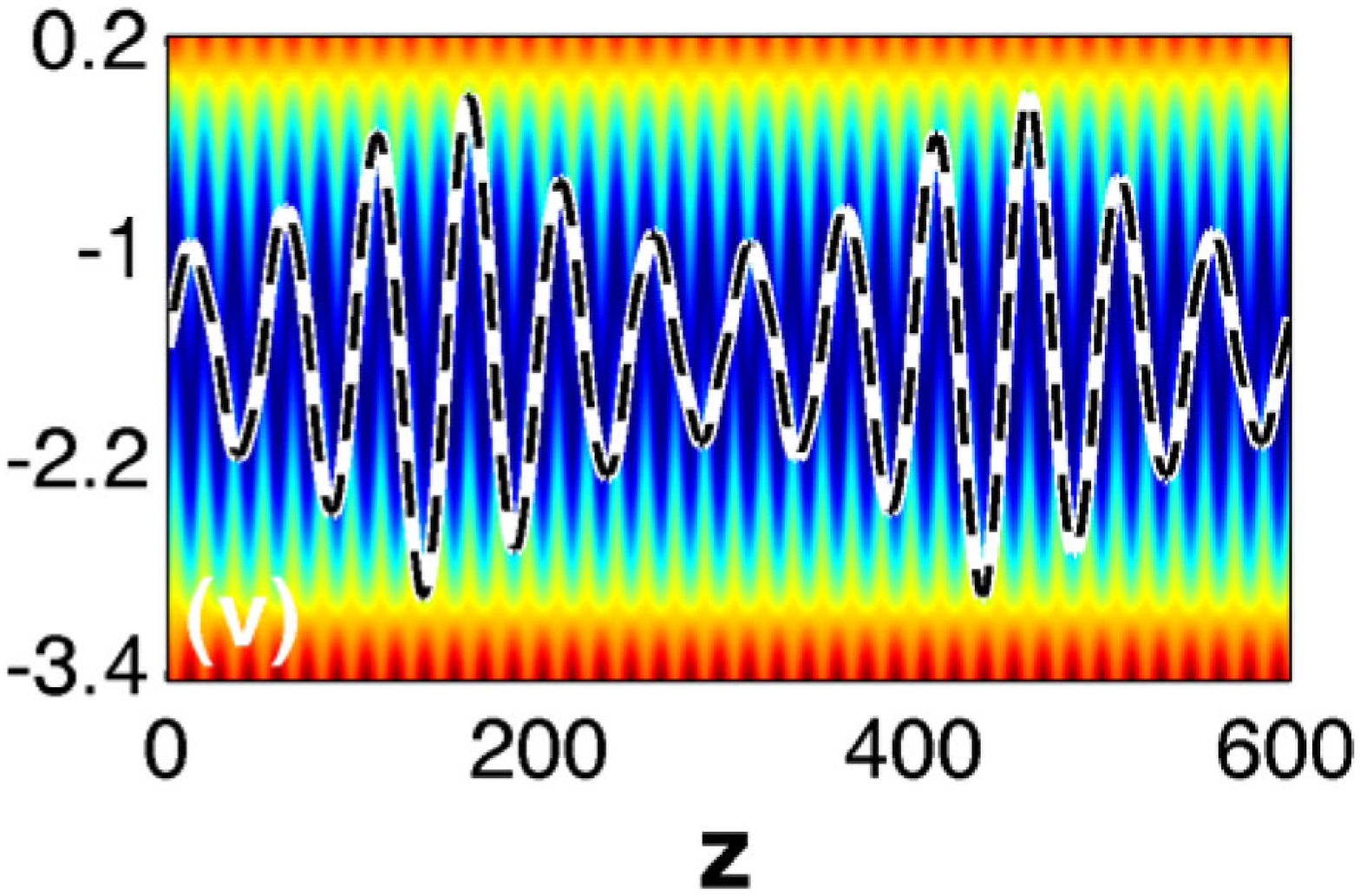}}}
        \caption{(Color online) Evolution of the soliton center in an WM lattice pattern with initial conditions taken from the corresponding points $(i)-(v)$ of Fig. 5(b). (a) point $(i)$ (upper, yellow curve): Trapping with periodic oscillations  for the soliton with the positive initial velocity, point $(ii)$ (lower, white curve): Dynamic switching for the soliton with the negative velocity. (b) point $(iii)$ (upper, yellow curve): Trapping with quasiperiodic oscillations for the soliton with positive initial velocity, point $(iv)$ (lower, white curve): Detrapped motion for the soliton with the negative velocity. (c) point $(v)$ (white curve): Beat oscillations. Thick white/yellow curves correspond to results from direct numerical simulations, dashed black curves from the effective particle model and solid black ones from direct simulations for the unmodulated lattice. The corresponding lattice pattern is shown in the background.}
   \end{center}
\end{figure}

\begin{figure}[p]
    \begin{center}
      \subfigure[]{\scalebox{0.5}{\includegraphics{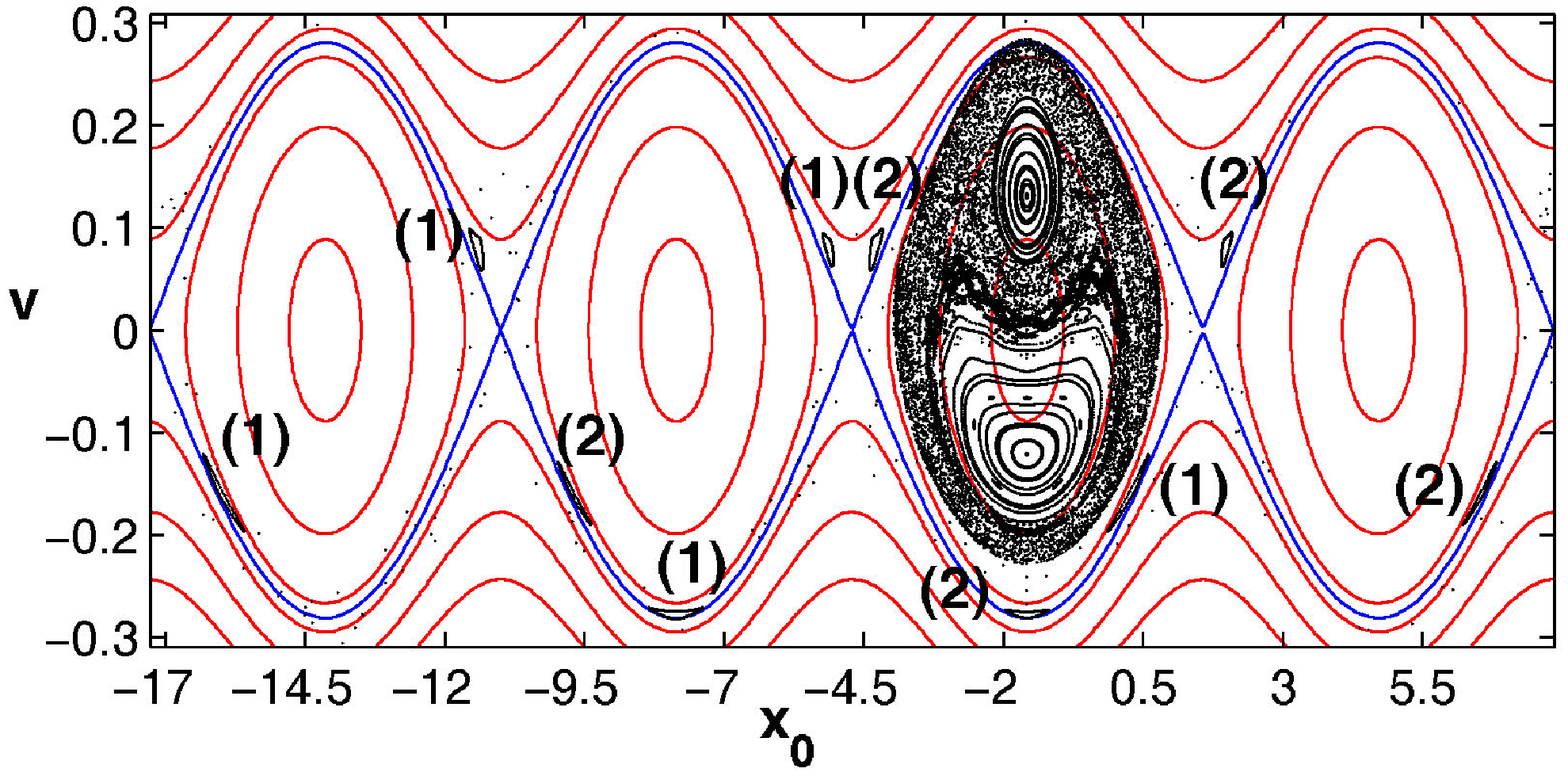}}}
      \subfigure[]{\scalebox{0.5}{\includegraphics{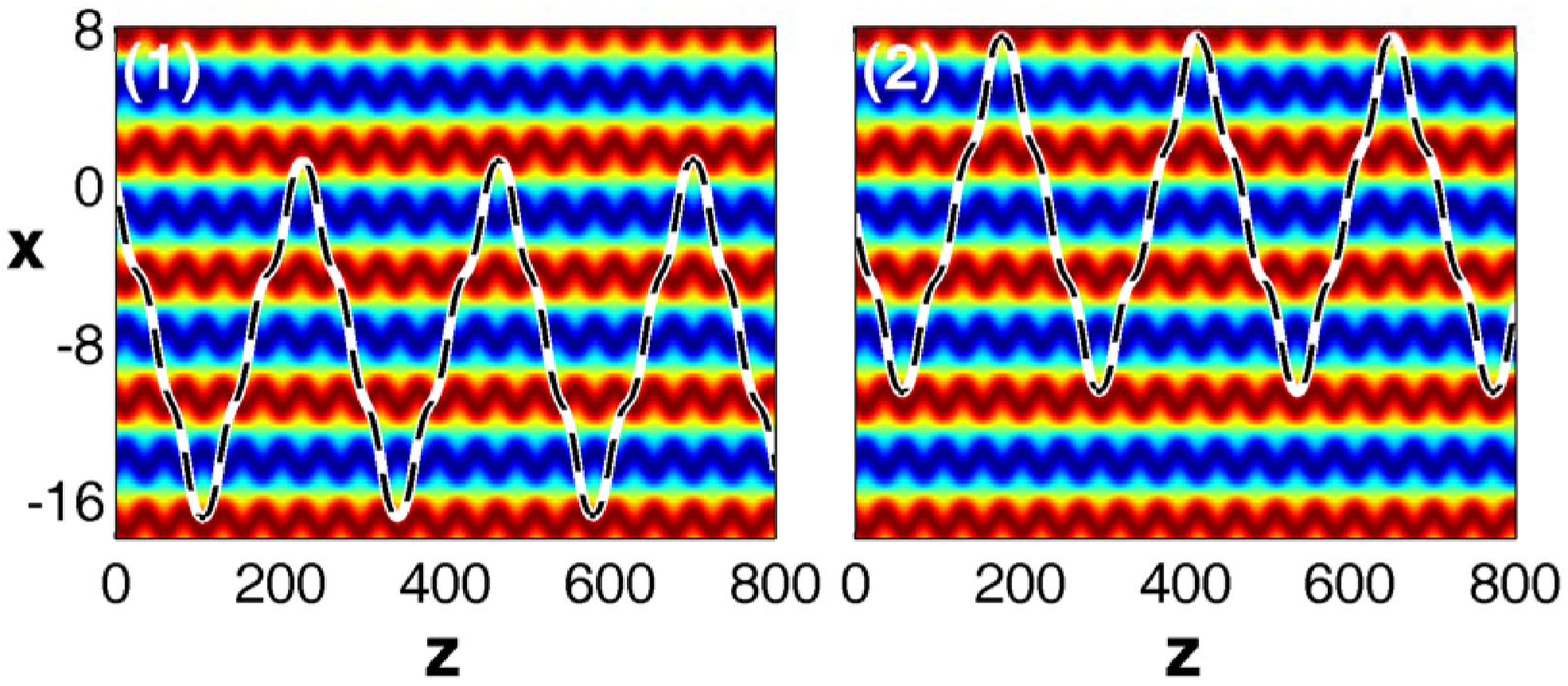}}}
      \caption{(Color online) (a) Poincare surface of section for a soliton with $\eta=5$ at a WM lattice potential of Fig. 1(d), superimposed on the phase space  of the unmodulated lattice. For illustration purposes, the surface of section is produced with initial values at one of the four periods of the phase space. (1) and (2) are two different families (belonging to different tori) of interconnected resonant islands. (b) Soliton center trajectories of each fammily with initial conditions (1) $(x_0,v)=(0.05,0.178)$ and (2) $(x_0,v)=(-\pi/2,-0.28)$. Thick white curves correspond to results obtained from direct numerical simulations, dashed black curves are obtained form the effective particle model.}
   \end{center}
\end{figure}


\begin{thebibliography}{99}
\bibitem{Chris} D.N. Christodoulides and R.I. Joseph, Opt. Lett. \textbf{13}, 794 (1988); 
H.S. Eisenberg, Y.Silberberg, R. Morandotti, and J.S. Aitchison, Phys. Rev. Lett. \textbf{81}, 3383 (1998);
J.W. Fleischer, M. Segev, N.K. Efremidis, and D.N. Christodoulides, Nature \textbf{422}, 147 (2003);
D. N. Christodoulides, F. Lederer, and Y. Silberberg, Nature \textbf{424}, 817 (2003).
\bibitem{mono_static} N.K. Efremidis and D.N. Christodoulides, Phys. Rev. A \textbf{67}, 063608 (2003); P.J.Y. Louis, E.A. Ostrovskaya, C.M. Savage, and Y.S. Kivshar, Phys. Rev. A \textbf{67}, 013602 (2003); D.E. Pelinovsky, A.A. Sukhorukov and Y.S. Kivshar, Phys. Rev. E \textbf{70}, 036618 (2004).
\bibitem{OpEx_08} Y. Kominis and K. Hizanidis, Optics Express \textbf{16}, 12124 (2008).
\bibitem{DuKe_86} D. H.  Dunlap, and V. M. Kenkre, Phys. Rev. B \textbf{34}, 3625 (1986).
\bibitem{Holthaus_92} M. Holthaus, Phys. Rev. Lett. \textbf{69}, 351 (1992).
\bibitem{DiSt_02} M. M. Dignam, and C. M. de Sterke, Phys. Rev. Lett. \textbf{88}, 046806 (2002).
\bibitem{MaMa_06} T. Mayteevarunyoo and B.A. Malomed, Phys. Rev. A \textbf{74}, 033616 (2006).
\bibitem{PoAlOst_08} D. Poletti, T.J. Alexander, E.A. Ostrovskaya, B. Li, and Yu.S. Kivshar, Phys. Rev. Lett. \textbf{101}, 150403 (2008); D. Poletti, E.A. Ostrovskaya, T.J. Alexander, B. Li, Y.S. Kivshar, Physica D \textbf{238}, 1338 (2009); J. Abdullaev, D. Poletti, E.A. Ostrovskaya,and Y.S. Kivshar, Phys. Rev. Lett. \textbf{105}, 090401 (2010). 
\bibitem{StLo_08} K. Staliunas, S. Longhi, Phys. Rev. A \textbf{78}, 33606(2008).
\bibitem{dm} H. S. Eisenberg, Y. Silberberg, R. Morandotti and J. S. Aitchison, Phys. Rev. Lett. \textbf{85}, 1863 (2000); 
M. J. Ablowitz and Z. H. Musslimani, Phys. Rev. Lett. \textbf{87}, 254102 (2001); A. Szameit, I.L. Garanovich, M. Heinrich, A. Minovich, F. Dreisow, A.A. Sukhorukov, T. Pertsch, D.N. Neshev, S. Nolte, W. Krolikowski, A. Tünnermann, A. Mitchell, and Y.S. Kivshar, Phys. Rev. A \textbf{78}, R031801 (2008).  
\bibitem{rabi} Y.V. Kartashov, V.A. Vysloukh and L. Torner, Phys. Rev. Lett. \textbf{99}, 233903 (2007); K.G. Makris, D.N. Christodoulides, O. Peleg, M. Segev and D. Kip, Optics Express \textbf{16}, 10309 (2008); K. Shandarova, C. E. Ruter, D. Kip, K. G. Makris, D. N. Christodoulides, O. Peleg, and M. Segev, Phys. Rev. Lett. \textbf{102}, 123905 (2009).
\bibitem{Kominis_steering} Y. Kominis and K. Hizanidis, J. Opt. Soc Am. B \textbf{21}, 562 (2004); J. Opt. Soc Am. B \textbf{22}, 1360 (2005).
\bibitem{Tsopelas_steering} I. Tsopelas, Y. Kominis and K. Hizanidis K, Phys. Rev. E \textbf{74}, 036613 (2006); Phys. Rev. E \textbf{76}, 046609 (2007).  
\bibitem{Other_dragging} Y.V. Kartashov, L. Torner and D.N. Christodoulides, Opt. Lett. \textbf{30}, 1378 (2005); C.R. Rosberg, I.L. Garanovich, A.A. Sukhorukov, D.N. Neshev, W. Krolikowski and Y.S. Kivshar, Opt. Lett. \textbf{31}, 1498 (2006); G. Assanto, L.A. Cisneros, A.A. Minzoni, B.D. Skuse, N.F. Smyth and A.L. Worthy, Phys. Rev. Lett. \textbf{104}, 053903 (2010).
\bibitem{wgd_curved} S. Longhi, M. Marangoni, M. Lobino, R. Ramponi, P. Laporta, E. Cianci and V. Foglietti, Phys. Rev. Lett. \textbf{96}, 243901 (2006); R. Iyer, J.S. Aitchison, J. Wan, M.M. Dignam and C. M. de Sterke, Optics Express \textbf{15}, 3212 (2007); I.L. Garanovich, A. Szameit, A.A. Sukhorukov, T. Pertsch, W. Krolikowski, S. Nolte, D. Neshev, A. Tuennermann and Y.S. Kivshar, Optics Express \textbf{15}, 9737 (2007); A. Szameit, I.L. Garanovich, M. Heinrich, A.A. Sukhorukov, F. Dreisow, T. Pertsch, S. Nolte, A. Tünnermann and Y.S. Kivshar, Nature Physics \textbf{5}, 271 (2009).
\bibitem{wgd_varwidth} K. Staliunas and C. Masoller, Optics Express \textbf{14}, 10669 (2006); S. Longhi and K. Staliunas, Opt. Commun. \textbf{281}, 4343 (2008).
\bibitem{var_refractive} A. Szameit, Y.V. Kartashov, F.Dreisow, M. Heinrich, T. Pertsch,S. Nolte, A. Tunnermann, V. A. Vysloukh, F. Lederer, and L. Torner, Phys. Rev. Lett. \textbf{102}, 153901 (2009); A. Szameit, Y.V. Kartashov, M. Heinrich, F. Dreisow, R. Keil, S. Nolte, A. Tünnermann, V.A. Vysloukh, F. Lederer and L. Torner, Opt. Lett. \textbf{34}, 2700 (2009); Y.V. Kartashov, A. Szameit, V.A. Vysloukh and L. Torner, Opt. Lett. \textbf{34}, 2906 (2009).
\bibitem{Kaup_Newell} D.J. Kaup and A.C. Newell, Proc. R. Soc. London, Ser. A \textbf{361}, 413 (1978).




\end{thebibliography}
\end{document}